**Learning from development of a third-party patient-oriented application using Australia's national personal health records system**

**Authors:**

Niranjan Bidargaddi [1], Michael Kidd[1]

1. School of Medicine, Flinders University, Adelaide, Australia

**Corresponding author**:

A/Prof Niranjan Bidargaddi

Personal Health Informatics, School of Medicine, Flinders University

Email: niranjan.bidargaddi@flinders.edu.au

# Learning from development of a third-party patient-oriented application using Australia's national personal health records system

**Abstract**

**Background:** Large-scale national level Personal Health Record (PHR) has been implemented in Australia. However, usability, data quality and poor functionalities have resulted in low utility affecting enrollment and participation rates by both patients and clinicians alike. Development of new applications deriving secondary utility of data can enhance use of PHR's but there is limited understanding on processes involved in development of third-party applications with nationally run PHRs.

**Methods:** Analysis of processes and regulatory requirements for developing applications of data from My Health Record, Australia's nationally run PHR and subsequently implementation of a patient oriented software application using data sourced from My Health Record. Synthesis of learning's from implementation experience into recommendations for improving third party application development with nationally run personal health records.

**Results:** The study revealed a nuanced understanding of different data types and quality of data in My Health Record and complexities associated with developing secondary use applications. Regulatory requirements associated with utilization of My Health Record data, restrictions on visualizations of data and processes of testing third-party applications were addressed to develop healthtimeline application, the first My Health Record data based open source application aimed at both patients and clinicians. Healthtimeline application translates Medicare claims records stored in myhealhthrecords into a clinically meaningful timeline visualization of historical data.

**Conclusions:** This study demonstrated processes involved in development of an application utilizing data in Australia nationally run PHR, My Health Record. The study identified several process, technical and regulatory barriers which if addressed have the potential to make My Health Record an attractive platform for many application developers resulting in a thriving ecosystem of applications.

**INTRODUCTION AND OBJECTIVES**

Across the OECD, implementations of eHealth environments consisting of integrated electronic health data, accessible by both health care providers and individuals are an emerging trend(1). In the USA the 2010 Electronic Health Record (EHR) Incentive program provided a impetus for health care providers to replace paper based health records with electronic records and make these data also electronically available to patients(2). Since American health care system is mostly run by private sector, this policy has resulted in a distributed network of eHealth implementations that are localized to health care agencies, with each agency providing portal based access to its members (3). Large-scale national eHealth implementations have been attempted in Australia(4) and several countries across Europe(1). Personal Health Records (PHRs) is the term used to describe both the health information and functionalities offered through such environments. The type of information available in PHR's varies from clinical documents, lab results to patient generated home monitoring data. Collection, sharing, exchange and self-management of information are the typical functionalities available in PHRs. Several factors, ranging from type of data sources available and functionalities, to enrollment and participation rates of health care providers patients influence the usefulness of PHR's.

Within implemented PHRs, the type of information stored and functionalities offered have varied substantially, with many being physician-oriented and lacking in patient-oriented functionalities(5). Rates of enrollment and participation with PHRs have also differed substantially between different implementation settings. In the USA, where PHR offering is through portals linked to individual health care agencies, relatively high participation rates are reported by members of some agencies. For example, 50% of Kaiser Permanante's Health Maintenance Organization plan members, one of the largest providers, were using its portal to access service in 2014(6). However, participation rates across portals are not uniform with several socioeconomic disparities observed(7). On the other hand, several nationally run PHR's hampered by severely low participation rates, such as the ones in UK, have been either withdrawn or scaled down(8). A notable exception is Australia's nationally run PHR, known as My Health Record (formerly known as personally controlled electronic health records) (4), incepted in 2012 with a reported enrollment of 9% of Australian population and over 5000 general practice health professionals (almost 75% of eligible practices) (9) over a 3 year period. It is anticipated with the introduction of an opt-out model for individual enrollment, which is currently trialed, problems associated with lower patient enrollment rates will be overcome. While this approach may overcome enrollment issues, usage and quality of participation would ultimately depend on the benefits individuals and providers alike derive from My Health Record. There have been several studies investigating end-user perspectives and experiences with My Health Record since 2012 (9-11). However, there is a critical gap in knowledge on actual utility of My Health Record as evident by lack of studies investigating My Health Record with participating individuals and providers that demonstrate outcomes or the development of new applications for enhancing utility and participation.

    Critical to the quality of My Health Record is the utility of its data and functionality in processes of disease prevention and management. For example, access to fragmented medical history in a single place is one of the main perceived benefit and utility of PHRs (11). In the case of My Health Record, the functionality for obtaining data from appropriate health data repositories exists, but a coherent approach to automatically source data from various repositories and create a collated health summary is missing (12). As a result My Health Record has limited utility for the purposes of obtaining fragmented medical history in one place and for the coordination of care until there is a completeness of records. In the absence of complete data, the better presentation and application of available data can still increase its usefulness. For example, it is well established that reminder functionalities linked to primary care data can result in cost-effective communications for prescription and appointments(13). However, within Australian context, most health care providers are less likely to utilize My Health Record as a

communication platform, since a vast majority already use practice management software's with advanced functionalities as well as most of the information intended to be made available through My Health Record (12).

Thus developing an understanding of different types of data available within My Health Record and exploring potential new applications is crucial to deriving value. For example, My Health Record interfaces with the Medicare claims database, which processes health interactions and drug prescriptions claims for Australians. The utility of Medicare claims data in health research has been widely demonstrated. The health research community has extracted this data directly from Medicare Australia and utilized it to derive patients' health service interaction and medications profile in a wide range of studies(14,15). However, these data have not been applied in clinical decision support. For example, a timeline visualization of medication and service use can reveal gaps in medications, typical service use patterns and their changes which provide clinical contexts relevant to assessment and treatments (16,17). Since the My Health Record platform contains real time Medicare claims data, it is highly suited for developing new applications of this data for clinical decision support and research. However, developing applications of My Health Record have been attempted and deemed difficult(18).

Developing an understanding of the processes of development of applications using My Health Record is important to realize benefits of already available and useful repositories of Medicare for participating patients and providers. This paper describes the experiences from successful development of a My Health Record software application, called *healthtimeline*. The healthtimeline application creates a novel timeline visualization of Medicare data in My Health Record and is aimed at both clinicians and patients (see Figure 4). The processes involved in developing interfaces with My Health Record and the use case conformance and testing requirements are described in detail. Key issues that impact development of My Health Record applications uncovered in the study and several strategies for developers along with recommendations for improvements to My Health Record are discussed.

**PROCESS OF DEVELOPING MYHEALTHRECORD APPLICATIONS**

**My Health Record**

My Health Record consists of a centralized infrastructure to verify, obtain and transfer health information of individuals from repositories holding health and clinical documents.  It is designed to interact with several distributed repositories, a few of which are established and managed centrally by the National Digital Health Agency (formerly known as National e-Health Transition Authority) and some are maintained by registered external organizations. For example, Medicare Australia's Medicare benefits and prescription claims database is one such linked repository. The central infrastructure is designed to query and identify the repositories that contain information about an individual patient using their unique identifier, known as Individual Health Identifier (IHI) and to collate available data on request from the repositories. Individuals access to view this data is only through a default portal offered by the central infrastructure. In 2016 the portal was revamped with an improved user interface (see Figure 1 & 2). Health professionals on the other hand can access the information either through a similar default portal or through a conformant third party clinical information system.

**INSERT FIGURE 1: My Health Record portal data view**

**INSERT FIGURE 2: Medicare data view in My Health Record portal**

*Data types available in My Health Record*

Health data stored in My Health Record is programmatically accessible in a format known as views. There are 8 different views pertaining to different types of available data.  The health data in each of these views is populated under different conditions with varying level of data completeness. Except for Medicare view, all views are populated by data sourced from various clinical information systems used by GPs, radiologists and hospitals. Also despite the availability of interfaces not all available data is automatically extracted from these sources and populated into corresponding views. As a result, it has been claimed health data in My Health Record is often incomplete making it unsuitable for patient care(19). A notable exception is data stored under Medicare view which contains details directly sourced from Medicare claims database. It consists of a list of Medicare Benefit Schedule (MBS) interventions and pharmaceutical benefits schedule (PBS) prescriptions received by the patient with variables describing the service type and date of occurrence. These records are complete and retrospective for 2 years from the date of activation of My Health Record account.

**Regulatory requirements for using My Health Record**

Developing a third-party patient application with interfaces to access My Health Record data is not straightforward. At the time of this study My Health Record proprietary portal was intended to be the only interface for patients to access data. Third party applications that were supported for integrations were clinical information systems aimed at health professionals. Furthermore, there are strict technical and intended use requirements. Third party software that integrates with My Health Record has to meet several conformance requirements that are detailed in next section. A conformant software could then be used to access data from My Health Record, but the access can be only for the purposes of providing 'health service' as defined in the 1988 Privacy Act, which is usually interpreted as a service that involves a clinician involvement. As a result, currently available software applications integrating with My Health Record are traditional clinical information system applications aimed at health professionals (20). So far no patient oriented third-party application utilizing My Health Record data that can satisfy provision of 'health care' requirements without requiring a clinician involvement has been developed.

**Technical conformance requirements for myhealthrecord applications**

Implementation of a third-party application that interfaces with My Health Record involves integration with two web services known as Healthcare Identifiers (HI)

Service, maintained by Medicare Australia and PCEHR service from Digital health Agency. In order to develop a My Health Record application, the first step is to register as an application developer with both these agencies after which access to a development environment and a test kit will be granted. The test kit contains sample data for testing along with descriptions of supported integration use cases.

The HI service is a service for verifying health care providers, health care organizations and individuals receiving health care in Australia using the unique 16-digit identifiers assigned under the auspices of Healthcare Identifiers Act legislation in 2010 (21). Third party applications have to verify the identity of their application users in the HI service in order to exchange data with My Health Record. This involves integrating the Business 2 Business Application Programmer Interfaces (API)'s functionality from HI service into the application and subsequently testing to ensure that the implementation is adherent to the pre-approved use cases. The HI implementation testing involves two phases, first demonstrating that application can exchange information with HI service, known as 'notice of connection' and secondly that the information exchange is consistent with approved use cases (also known as 'conformance testing'). Third-party applications that can successfully verify identify of their users in HI service are eligible to read and write My Health Record data using API functionalities offered by PCEHR service. The first step involves calling PCEHR APIs to check if My Health Record actually exists for the HI verified individual (as this was a opt-in system).  If valid records exist, subsequent steps involve invoking various APIs that provide functionalities for reading or writing different types of medical records in My Health Record. In this study only reading data from PCEHR service was explored which was done using 'getview' API. As with HI service, third party applicatios that read data from PCEHR service have to adhere to rendering guidelines while displaying this information. The testing involves again a notice of connection and conformance testing to demonstrate that functions of PCEHR record validation and 'getview' API rendering guidelines are correct.

**Accessing My Health Record production server**

A developed application has to meet the regulatory requirements for use, obtain clearance for the four tests described above, each with different teams, before details for production level access are granted. There is a fee associated with HI conformance test, as it needs to be carried out by an Australian National Association of Testing Authorities (NATA) accredited test vendor. Once the tests are passed and notified to Department of Health, access to My Health Record production server through the developed application is granted.

**RESULTS: IMPLEMENTATION OF A MY HEALTH RECORD APPLICATION**

**Healthtimeline application design**

Healthtimeline is a responsive web application that visualizes My Health Record Medicare view data in form of a timeline view. It has both patient and clinician interfaces, but is designed to be a standalone application accessible by registered patients even when there is no interaction with health professionals involved. Thus, it is the first patient oriented third party application interfacing with My Health Record. The development of healthtimeline application involved programming interfacing with HI and PCEHR services, applying timeline visualizations on Medicare data and finally testing conformance requirements. The application was developed uses open source development and hosting tools. It is developed in Java Enterprise Edition using JBOSS Seam and Hibernate frameworks. The database is created in postgresql. JBOSS Application server and APACHE web servers are required to run the compiled application. HL7 SOAP protocol is used for integration with HI and PCEHR web services.

The Personal Health Informatics team from Flinders University was registered as a software developer with Medicare Australia and Digital Health Agency (formerly NEHTA) to gain access to developer and testing environments. The project was conducted over 12 months, with significant part of this time spent communicating with the Digital Health Agency and Department of Health staff clarifying the processes for interfacing and testing. The application was developed to meet the provision of 'health care' as a standalone application and satisfy conformance requirements through processes and components outlined below (Figure 3).

**INSERT Figure 3: Healthtimeline application architecture**

*Registration and login process*

The application has an online option for registration either as a patient or a clinician user type. Any individuals can register with the application as a patient user. This step involves setting up a username, which is a mobile phone number and password as well as providing details required for the purposes of verifying with Health Identifier (HI) Service. Individuals can either provide their 16-digit IHI number, or alternatively provide other details (first name, last name, date of birth, Gender and Medicare number). The application sends this information to HI service using IHI inquiry via B2B (reference) API for verification and upon receiving a valid user confirmation user record is created in the healthtimeline application's database.

The registration page for clinician type user also involves entering mobile number and password similar to the patient user type. The registration page also has the capacity to collect 16-digit Healthcare provider Identifier (HPI-I), another unique identifier used by My Health Record for verifying health professionals. The application will send this information again to HI service via Health Care provider directory search API for verification purposes. However, discussions with Digital Health Agency technical staff revealed most health professionals will not know their HPI-I and that the collection and verification of this detail is compulsory if the use case involved a clinician creating a patient user account without patient's prior knowledge. The current release of the healthtimeline software only allows a clinician user type to access a patient account type that has already been created by individuals through processes detailed above.

*Applying timeline visualizations on My Health Record data*

The healthtimeline application applies a visually rich interactive timeline visualization on Medicare view data, implemented using vis.js ([http://visjs.org/](http://visjs.org/)) and timeline.js, two rich open source browser based visualization libraries. The timeline visualization comprises of displaying each MBS and PBS claim stored in Medicare view as an event against a time scale on horizontal axis time scale (Figure 4). The events are shown at the corresponding start date position displayed on the horizontal axis and the width of the event box can be adjusted to corresponding end date position if it exists. Each event is associated with a group and all events with the same group are shown in the same row or adjacent rows, as a box and the common value of their group property is used as a label at the left edge of the timeline. A simple taxonomy was developed to create a hierarchical grouping of PBS prescriptions and MBS interventions such that they are clinically meaningful and that variables required for assigning them into the groups were available in the Medicare view data. The MBS interventions were grouped into four categories; namely, GP's, specialists, imaging and pathology and PBS prescriptions were grouped into generic medication classes. Within each MBS group there were two subgroups based on if the service was provided in hospital or out of hospital and within each subgroup further levels based on nature of service. The group labels are displayed at the left side of the timeline visualization. Each MBS intervention and PBS prescription contained in Medicare view is then displayed as an event on timeline in the row of group it belongs to. The time scale on the horizontal axis can be adjusted with zooming options, which dynamically scales the visualized data. For each event box additional contextual information, some in the box or on the top in a banner is displayed. In the case of MBS interventions contextual information can be description of the claim such as "Optometrist visit" and the details of the service provider as well as if the service was provided "in hospital" or "out of hospital". For PBS prescriptions the contextual information includes prescription name, dispensed date and number of medications supplied. In addition to the above visualization, a simple notes entry page was also included. This was designed to facilitate patients and health professionals interpret and record insights or alternatively enable patients to record on topics/questions they would like to discuss further.

**INSERT FIGURE 4: Timeline visualization of Medicare data**

Displaying timeline visualization of Medicare data obtained from My Health Record required overcoming several standards related challenges. In order to meet the software conformance standards health data had to be displayed in a predefined format and style guidelines set by Digital Health Agency (22). The guideline specifies 67 different requirements covering font size, format and structure and not all of these were applicable to the formats and style of timeline visualizations. For example, in the time visualization category taxonomies derived from the raw claims data are displayed, but in order to be conformant all the raw data has to be displayed. The information also needs to be displayed in the same order it appears in the clinical document but in the timeline view they are displayed as events. In order to meet conformance requirement, the application had to implement a conformant approach to display newly fetched data from My Health Record.  Subject to implementation of a conformant approach no restrictions applied on alternative secondary visualizations.  As such two different data visualizations were implemented on the same Medicare data, rendering using an approved style sheet provided by NEHTA for the purposes of meeting conformance as well as simultaneously creating the alternative timeline visualizations described above in a separate tab.

*Process of accessing data from My Health Record*

Healthtimeline application is designed with processes to access and share data sourced from My Health Record with informed consent. It can only obtain My Health Record data, for an individual who has registered as a patient user type in the application, after electronically consenting to the terms of use and privacy policy that describe how the information is used and where it will be hosted. A patient user type after successful login can see their profile details and view their My Health Record data in the healthtimeline visualization. In order to create the visualizations the application has to first fetch the data from My Health Record, store them in the database against the patient user type's records and subsequently create visualizations. The process for fetching data involves first verifying if the individual with the IHI number has a myhealthrecord by calling *checkifPCEHRexists* API, and upon successful confirmation gain access to myhealthrecord database. Since My Health Record was an opt-in system at the time of this project, this step checks if the user has activated their My Health Record. After receiving successful confirmation that the user has an active My Health Record, the application then fetches the health data using Getview API's.

A registered clinician user has to be connected to a patient user before they can view their record. The application can be configured to allow a clinician user type either to be automatically connected to all or selectively connected to registered patient users. If a clinician is not designated to have automatic connection with all individuals, they can use the search option by entering patient name, date of birth, gender and Medicare number to find matching patients and connect. Furthermore, a patient user can view a list of clinicians registered as a clinician user type on the application and they can control which of these clinicians has access to their data by using a connect or disconnect button. The application is also programmed to extract latest records from My Health Record, when a patient profile is opened, either by a patient user type or a connected clinician user type. The host of the application can download de-identified dataset of participants for research purposes.

**Testing process**

The developed application under went four different tests within the vendor environment for Health Identifier (HI) and PCEHR service, assisted by staff at Digital Health Agency. HI Notice of Connection (NOC) test involved taking a screenshot and log file based on test cases and were approved by testers at the Department of Health. After completing the NOC test, HI conformance test was conducted by IV&V Australia, an accredited tester, at a cost of $10,000. The process involved first creating test cases and subsequently remotely assessing the using sample data. The Department of Health and Ageing issued an approval letter with details to gain access to production HI service, subsequent to passing the conformance tests. The PCEHR NOC test was carried out Ventura Inc. and there was no cost associated with this test. Testing involved verifying appropriate warning and alerts are displayed in the application for incorrect individual details included in sample data. A tester from Accentura remotely monitored the application while the developer tested different patient and clinician scenarios. Finally, the PCEHR conformance test, which involved assessment of rendering guidelines, was done through a self-assessment form. The

results of the self-assessment and a completed PCEHR vendor declaration conformance form along with the HI production access letter received earlier were then sent to Department of Health following which a letter with details for access to production server of PCEHR was issued.

Sample datasets in test kit were mainly aimed at testing the authentication and data access process and not for utilization needs. They did not contain longitudinal records nor did they sufficient number of test cases required for refining the visualization categories. A separate data set sourced from a different study that has fields contained in Medicare view was used to develop the timeline analytics(23). The source codes of the certified application are available at https://bitbucket.org/pcehr/flinders.git.

**Requirements for hosting healthtimeline application**

Administering healthtimeline application involves hosting the application on a server and meeting eligibility to be a host organization by Digital Health Agency. The host-organization needs to adopt a My Health Record use policy and register to be a My Health Record "participating organization" and apply to obtain a Health provider identifier. The purpose of My Health Record use policy is to ensure that the organizations are accessing data through conformant software for providing health care only. In this situation registered organization will be accessing My Health Record data through the healthtimeline application for providing personal health data insights as a service to registered patients and clinician user types. On approval a HPI-O number, which is 16-digit unique organization number verifiable by HI service and a National Authentication Service for Health Public Key Infrastructure Certificate, which is a digital certificate for activating the conformant software, are provided. Both these details need to be keyed into the healthtimeline application and only then can the application can make connections with live My Health Record. Individuals signing up as clinician and patient user types have to read and consent to terms of using healthtimeline service and privacy policy of healthtimeline application, which were developed in consultation with Flinders University legal team. It contains information on their obligations and outlines how information sourced from My Health Record are used and managed. For the purposes of this study South Australian Health and Medical Research Institute was registered as a 'participating organization', and a demonstration instance of the healthtimeline application is hosted on Nectar, a cloud server infrastructure available for Australian University researchers under the web address www.healthtimeline.co.

**DISCUSSION OF ISSUES**

The preceding sections described steps involved in developing applications for My Health Record, Australia's nationally run PHR and described the successful development of a healthtimeline application utilizing Medicare data from My Health Record. Several challenges were encountered during development that can be grouped in four categories, namely a) regulations related to use of My Health Record in applications, b) type of applications of My Health Record data c) issues related to data processing and d) regulations related to testing. These are discussed below.

**Regulations related to use of My Health Record data in applications**

A major challenge with development of secondary applications arises from both the way in which new applications can interface with My Health Record and the requirements surrounding how this information should be used. My Health Record has a default patient and provider portal for data access. Access to data stored in My Health Record is either through these default portals or through third party conformant software's interfacing with provider portal. An entity eligible for recognition as a 'health service provider' is required for operating the conformant software. Furthermore, the PCEHR 2012 legislation (24) stipulates eligible entities should utilize My Health Record data only for the purposes of providing 'health service' which is defined in the Privacy Act 1988 as *"(a) an activity performed in relation to an individual that is intended or claimed (expressly or otherwise) by the individual or the person performing it:*

*(i) to assess, record, maintain or improve the individual's health; or*
*(ii) to diagnose the individual's illness or disability; or*
*(iii) to treat the individual's illness or disability or suspected illness or disability; or*

*(b) the dispensing on prescription of a drug or medicinal preparation by a pharmacist*.". All existing third-party conformant software satisfy these criteria, as they are all purely health-professional oriented. However, this definition is ambiguous about standalone patient-oriented self-management health care applications. Patient-oriented standalone applications such as the healthtimeline, which apply visualization and analytics on data and provide feedback, with or without health professional involvement are a new category of "health service" widely known as online or internet personal health interventions. Standalone applications meeting these criteria are prominent and shown to be effective in health, for example in treatment of depression(25) or cardiovascular rehabilitation (26). However, so far there hasn't been any standalone patient-oriented health application developed using My Health Record, possibly due to the lack of clarity around if an application itself can be treated as a 'health service' provider. Furthermore, My Health Record has adopted the health service definition from 1988's Privacy Act when digital health applications were not envisaged. One solution is to consider the organizations creating and hosting standalone patient-oriented applications of My Health Record, as a health service provider, which will result in opportunities for creating new application using My Health Record data.

Another challenge is related to restriction on the use of My Health Record data for health care purposes only, as some secondary use cases such as research and analysis might not meet the criteria of the 2012 Australian PCEHR legislation. However, the restriction on using data for healthcare purposes only applies while retrieving data from My Health Record. Once the information is lawfully obtained from My Health Record, local terms of use and privacy policy within the application can be applied for subsequent utilization of downloaded data to support new use cases. This approach helps overcome challenges with creating new use cases for the healthththtimeline such as providing treatment decision support, data linkage endeavors or recruitment for clinical trials(27).

**My Health Record data application use case considerations**

Development of an application of My Health Record with new use cases also requires considerations on the type of information that will be used towards these use cases. The quality of information pertaining to different data views in My Health Record differs substantially. The completeness of information under each view is dependent on the approach used for data population. Views containing automatically sourced health data stored have complete data and are thus useful for innovative analytics and clinical trial use cases. For example, data in Medicare view is automatically populated from Medicare Benefits scheme and Pharmaceutical Benefits scheme claims repositories, which record details of services and prescription claims made by individuals against these schemes. Since the Medicare processes claims as and when they happen, Medicare data view is complete and provides a complete trajectory of individuals. The utility of this data is also well established in understanding patterns of service use and medication adherence (14). Applications that are focused on analytics and less on data collection processes would benefit from this data repository. On the other hand, data in other views such as clinical documents and lab results are not automatically populated and the processes for sourcing data are not uniform. For instance, in the case of views that require data sourced from GP records, individuals have to discuss and request their GP's to upload the information into these views (28). Furthermore, the data population strategy for several of the other views like radiology or lab results which do not hold any data appears to be not fully developed and clear. Thus, developing new applications using data from views that are not automatically sourced, may need to also factor in development of strategies for data collection.

**Data processing challenges**

Utilizing My Health Record data for analytics applications presents several data processing challenges. The first relates to the way data is described. Analogous to shared file repositories, such as Dropbox$^{TM}$, My Health Record is a collection of clinical documents sourced from different repositories with interpretation and use of information in the documents left to individuals using it. The data in different views is characterised by clinical terminologies and jargons specific to repositories from where they are sourced. This requires understanding and developing approaches to translate these terminologies suitable for new applications. For example, data in Medicare view, which is sourced from the claims repositories, contains service item numbers and prescription codes of claims made under Medicare. In this study, we devised a simple taxonomy that maps the item type and prescription codes into clinical categories. The clinical categories were based on service provider type or prescriptions for each disorder. The mapping involved creating a parser that groups item number associated with similar service provider and grouping of prescriptions associated with same disorders such as antidepressants.  Development of standardized taxonomies that described the data from a utility perspective would be critical to accelerating development of applications using of data from various views. Similar efforts to reduce complexity in data using taxonomies have been attempted in apps and wearable (29) and online marketing space (30) .

The second challenge of using My Health Record data, particularly in software applications, relates to visualization and rendering of data sourced from My Health Record.  Understandably to ensure safety and quality, My Health Record applications are required to adopt rendering guidelines while displaying health data. However, these guidelines are extremely specific as to how the information is displayed from font size, colour, format and content which makes it nearly impossible to implement visualizations or new ways of presenting data. For example, the timeline visualization developed in this study would not meet the rendering guidelines for displaying Medicare view data as it uses charts instead of free text. One way to overcome this challenge involved implementing both the default rendering and the new visualization and obtaining conformance with the default rendering default-rendering view. However, this is an inefficient process and instead relaxing the rendering guidelines would be preferable so that the focus can be on trying new visualizations of data.

**Regulations related to development and testing**

The ease of integration with My Health Record data is an important factor in developing secondary applications. Development with My Health Record involves integration with 2 separate services, with the first one being responsible for user identity verification is managed by Medicare and the other one for data access from Digital Health Agency. These interrelated services are coordinated by two different organizations, thus creating a complex workflow for application developers. Furthermore, a developed application is to be tested separately for an active connection and conformance to standards for each of these services. This is a complex process that not only involves testing several elaborate scenarios even if they are not the scope of the developed application but also conducting these with four different teams. In this project, the time and resource allocated for meeting conformance requirements were substantially higher compared to the actual use cases that were developed. This process might explain the substantially low number of applications developed using My Health Record so far (20).

Significant support is offered by the team at the Digital Health Agency to help developers navigate the process, but instead streamlining the testing process could actually reduce the time and costs for both application developers and the Agency. Thus, adopting streamlined integration and testing processes within My Health Record, albeit like the ones used in consumer health apps and wearable space, can create an impetus for a ecosystem of applications with innovative use cases of health data to emerge which ultimately will improve engagement and generate value. Finally, improving the size and richness of test data provided to application developers such that it reflects complete records of several patients can also assist in helping development of new applications.

**CONCLUSIONS**

My Health Record, Australia's national PHR maintains several views of health data sourced from different repositories with varying levels of data completeness in each view. This study demonstrates processes involved in developing applications of My Health Record data for secondary purposes. It developed healthtimeline, the first My Health Record data application aimed at both patients and clinicians, which overcomes regulatory constraints related to use and display of data. Healthtimeline application translates administratively processed prescription and Medicare benefits claims records stored in My Health Record into events of clinically meaningfully categories and creates a timeline visualization of events revealing patterns and insights for both clinicians and patients. The resulting application is developed to be registered as a stand alone 'health service' that can be accessed directly by the patients with or without a treating clinician and satisfy the requirements of 2012 Australian PCEHR legislation. The source code of the developed application is available for development of future patient-oriented applications. Addressing technical and regulatory recommendations made in the paper has the potential to position My Health Record as a platform with a thriving ecosystem of health applications in Australia.

**Acknowledgments**

**Contributors**

**Conflicts of Interest**

None to Declare